\documentclass{jkas}

\def\year{2018} 
\def\month{April} 
\def\volume{00} 
\def\issue{0} 
\def\beginpage{1} 
\def\endpage{99} 
\def\received{---} 
\def\accepted{---} 
\date{Received \received; accepted \accepted}

\title{
A Unified Spectral Model for Accreting Neutron Stars
}

\author[1]{Ayan Bhattacharjee}
\author[2]{Sandip K. Chakrabarti}

\affil[1]{S. N. Bose National Centre for Basic Sciences, Salt Lake, Kol: 700106, India; \email{ayan12@bose.res.in}}
\affil[2]{Indian Center for Space Physics, 43 Chalantika, Garia St. Road, Kol: 700084, India; \email{sandipchakrabarti9@gmail.com}}

\def\corrauthor{
	Ayan Bhattacharjee
}

\def\runningauthor{
Bhattacharjee and Chakrabarti
}

\def\runningtitle{
Spectral Model for Neutron Stars
}

\def\keywords{
radiation: dynamics --- stars: neutron --- accretion, accretion discs --- shock waves
}

\def\abstracttext{
Recent studies show that accretion flows around weakly magnetic Neutron Stars undergo multiple shocks, before reaching the surface of the star, which contribute to the spectral and timing variabilities observed in the X-Rays. Here, we report, for the first time, the spectral properties of a unified model of shocked accretion flows around Neutron Stars, based on the Two-Component Advective Flow paradigm. We compare our theoretical results with the X-Ray spectral features of Z and Atoll sources, across different states. We also fit RXTE/PCA spectra of Sco X-1 and 4U 1705-44 to show the potential application of this new model.
}

\begin{document}
\jkashead 


\section{Introduction}
Accretion flow around a black hole or a neutron star emits high energy radiations with varying spectral and temporal properties. Observed temporal variations point to the existence of a mechanism, dictated by the flow dynamics and not by the stellar surface or magnetic fields, that is common in both types of compact objects (Psaltis, Belloni, \& van der Klis, 1999; Belloni, Psaltis, \& van der Klis, 2002; Mauche, 2002). Spectral energy distributions of multiple sources indicate that the Comptonization process, the dominant mechanism for changing states in X-ray, takes place inside the flow which has similar physical properties in both the objects (Barret, 2001; Barret \& Olive, 2002; Paizis et al. 2006; Mendez 2006). The two-component advective flows (TCAF, Chakrabarti 1995; Chakrabarti \& Titarchuk 1995; Chakrabarti 1997) satisfy an array of such required properties and can thus be treated as a possible common solution when the magnetic field is low (Chakrabarti 2016; Bhattacharjee, \& Chakrabarti, 2017; Bhattacharjee, 2018; Bhattacharjee \& Chakrabarti, 2019). A TCAF based spectral model for the neutron stars to satisfactorily explain the observed spectral variations is therefore essential. Here we show that additional three physical parameters to existing TCAF solution hitherto applicable only for black holes, one can fit the spectra of accreting neutron stars and reproduce intricate features of two broad classes of sources (Van der Klis 1989), viz., Z and Atoll. We include soft photons from both the Keplerian disk (Shakura \& Sunyaev 1973) and the normal boundary layer (Chakrabarti \& Sahu, 1997; Bhattacharjee \& Chakrabarti, 2017; 2019) of the star and compute the combined Comptonized spectra. Existing models used multiple phenomenological parameters to fit the spectra and provided limited insight into the physical processes and the variation of the geometry of the accretion flow (Bhattacharjee 2018). However, TCAF-based model maintains its robust geometrical configuration away from the compact objects while accommodating boundary layer(s) in the accreting matter around both the black holes and neutron stars. To our knowledge, ours is the first physical model which tries to unify the accretion flow solutions in presence of these compact objects.
\begin{figure}[h]
\begin{center}
\includegraphics[height=4.5cm, width=8.0cm]{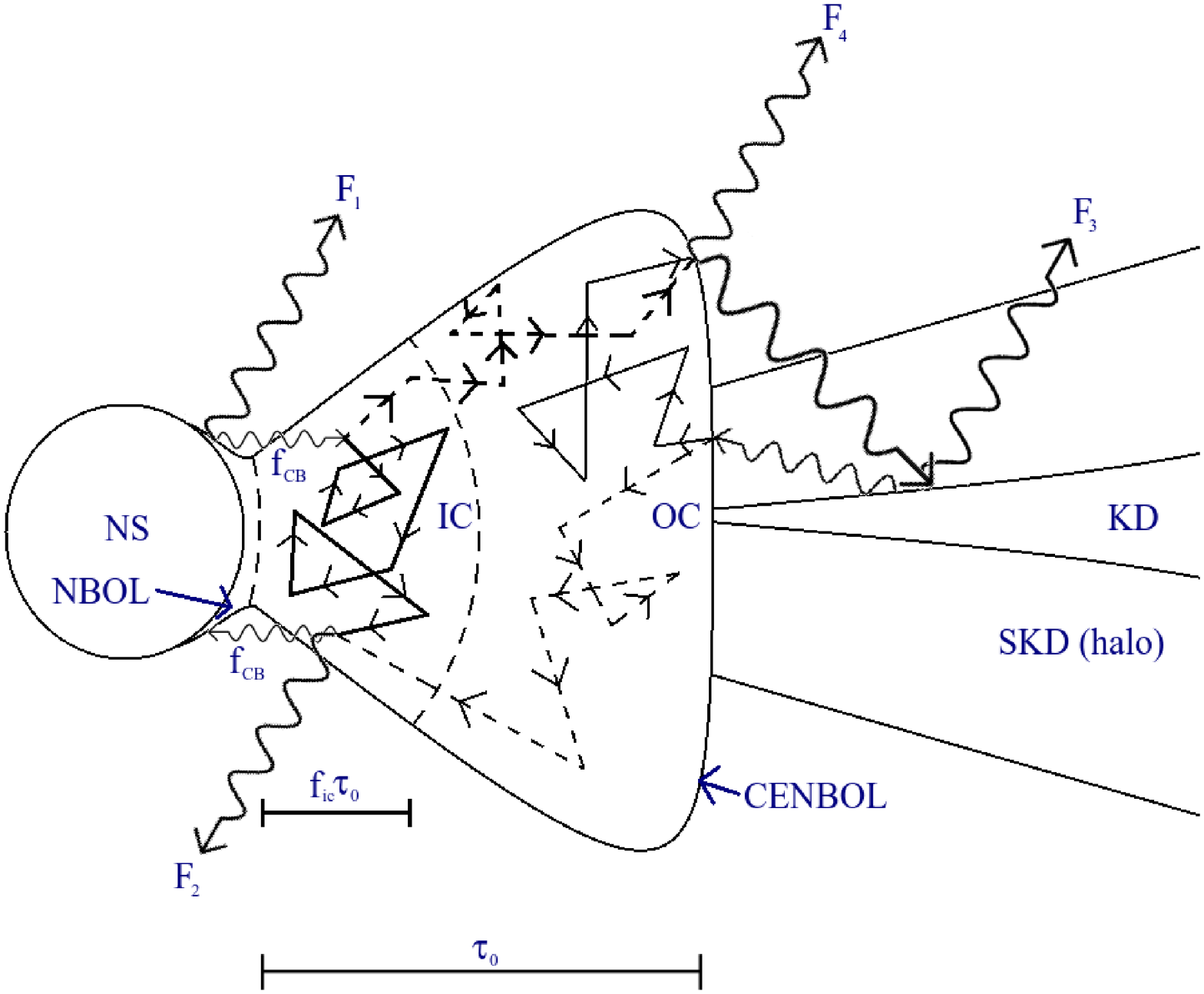}
\caption{\footnotesize{A schematic of the TANS model.}} 
\end{center}
\end{figure}

\section{Method}
The TCAF solution (Chakrabarti, 1989; 1990; 1995; Chakrabarti \& Titarchuk 1995; Chakrabarti 1997) has been used to explain to spectral and timing properties of stellar mass black holes in a self-consistent manner for multiple transient, persistent and class variable sources (Debnath et al. 2014; Bhattacharjee et al. 2017; Banerjee et al. 2019; Banerjee et al. 2020). The paradigm was also applied to Neutron Stars with some modifications using Monte Carlo simulations for the thermal Comptonization process (Bhattacharjee \& Chakrabarti, 2017). We use the empirical results of Bhattacharjee and Chakrabarti (2017, 2019) to modify the spectral code of TCAF for Neutron Stars by adding three new physical parameters (parameters 5-7 below) which are required to capture physics around ever-present boundary layer in Neutron Stars.
\subsection{Description of Model}
We use the following parameters in the spectral code TCAF Around Neutron Stars or TANS (Figure 1): 1. $\dot{m}_d$: disk (or Keplerian) accretion rate (in $\dot{M}_{Edd}$), 2. $\dot{m}_h$: halo (or sub-Keplerian) accretion rate (in $\dot{M}_{Edd}$), 3. $X_s$: location of the outer shock (=truncation radius of the disk) (in $r_S$), 4. $R_{comp}$: compression ratio of the outer shock, 5. $T_{NBOL}$: temperature of the Normal BOundary Layer or NBOL (in keV), 6. $f_{CB}$: fraction of NBOL photons intercepted by CENBOL (and vice-versa), 7. $f_{ic}$: transparency factor of the inner cloud. If the optical depth of the entire CENBOL is $\tau_0$, inner cloud transparent to the star till $\tau$, $f_{ic}\tau_0$). Here $\dot{M}_{Edd}$ is the Eddington rate for the Neutron Star and $r_S=2GM_{NS}/c^2$, is its Schwarzchild radius. For the present version of the code we kept the mass of the star constant at $M_{NS}=1.4 M_{\odot}$. A single suitable normalization (N) is required to match observed spectra with the theoretical one across spectral states. In Fig. 1, we show schematically how the seed soft photons from the NBOL and the KD are Comptonized by CENBOL electrons and are emitted to the observers. The expected TCAF configuration around a Neutron star has two components of accretion, namely, Keplerian Disk (KD) and sub-Keplerian or halo (SKD). The CENtrifugal pressure dominated BOundary Layer (CENBOL), which is the truncation radius of KD, serves as the Comptonizing region. The subsonic post-shock matter inside CENBOL hits the Neutron Star surface after passing through the inner shock which is the outer edge of the Normal BOundary Layer (NBOL). Seed photons from KD and NBOL are intercepted by CENBOL and are reprocessed to higher energy which reach the observers either directly or after getting reflected by the KD or NBOL. NBOL and KD dominantly interact with different regions, namely the Inner Cloud (IC) and the Outer Cloud (OC) respectively, of CENBOL due to proximity effects (Bhattacharjee and Chakrabarti 2017). The separation is characterized by a parameter $f_{ic}$, where inner cloud extends up to optical depth of $f_{ic}\tau_{0}$ from the NBOL. Fraction of photons received by CENBOL from NBOL and vice-versa is determined by $f_{CB}$. The final spectrum is obtained by combining the four fluxes: $F_1$, $F_2$, $F_3$, and $F_4$. Here, $F_1$ is the effective flux from NBOL (Blackbody+Thermally reprocessed), $F_2$ is Comptonized flux from the inner cloud, $F_3$ is the Comptonized flux from the outer cloud, and $F_4$ is the effective flux from KD (multicolour Blackbody+Thermally reprocessed). 

\section{Results}
\subsection{Computation of Spectra}
In Fig. 2, we give examples of our computation of a few typical spectra of the generalized flow configuration TANS as the parameters are varied in order to impress that our model is capable of reproducing observed spectra in Horizontal Branch (HB), Normal Branch (NB; Bottom Normal Branch: bot NB), and Flaring Branch (Bottom: bot FB; Middle: mid FB; Top: top FB). The computational parameters for all the cases are listed in Table 1. In Fig. 3 we show that the Island Branch (IS), Lower Left Banana Branch (LLB), Lower Banana Branch (LB), and Upper Banana Branch (UB) spectra can be generated by our model as well. For the set of parameters for these cases, see Table 2.

\begin{table}
\centering
\caption{\footnotesize{Parameters for Z type spectra.}}
\tiny{
\begin{tabular}{c c c c c c c c c}
\hline
\hline
No. & Class & $\dot{m}_d$ & $\dot{m}_h$ & $X_s$ & $T_{NBOL}$ & $f_{ic}$ & $f_{CB}$ \\~~\\
\hline
S1 & HB & 1.5 & 2.0 & 15.0 & 0.60 & 0.10 & 0.500 \\
S2 & NB & 1.5 & 1.5 & 15.0 & 0.60 & 0.10 & 0.500 \\
S3 & bot NB & 2.0 & 1.0 & 15.0 & 0.60 & 0.50 & 0.500 \\
S4 & bot FB & 2.5 & 0.5 & 15.0 & 0.60 & 0.50 & 0.250 \\
S5 & mid FB & 1.5 & 1.5 & 15.0 & 0.60 & 0.25 & 0.250 \\
S5.2 & mid FB & 1.5 & 2.5 & 30.0 & 1.00 & 0.25 & 0.250 \\
S5.3 & mid FB & 1.5 & 2.5 & 30.0 & 1.00 & 0.10 & 0.125 \\
S6 & top FB & 0.5 & 2.5 & 15.0 & 1.00 & 0.05 & 0.125 \\
\hline
\hline
\end{tabular}
}
\end{table}

\begin{figure}[h]
\begin{center}
\includegraphics[width=4.5cm, height=4.5cm]{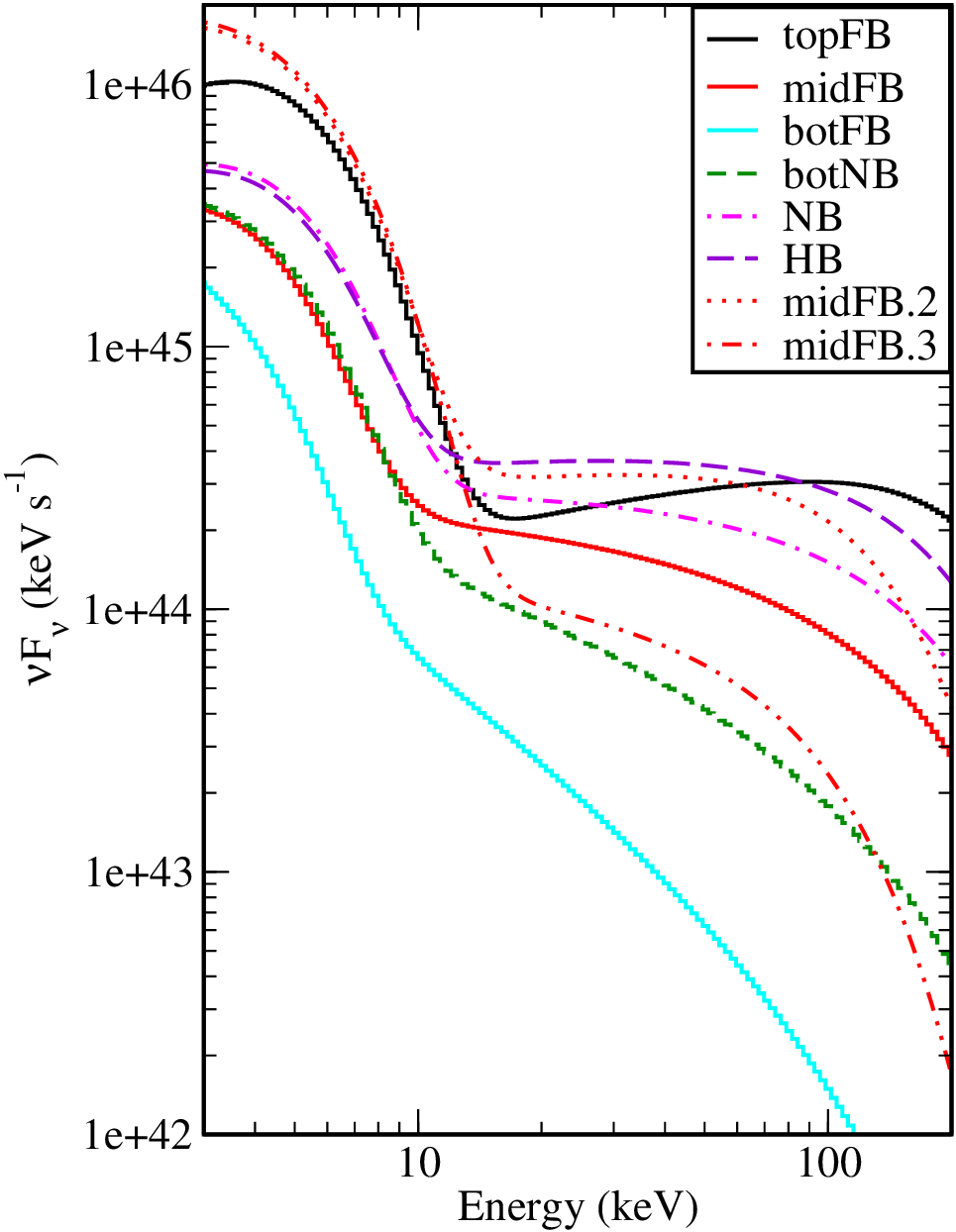}
	\caption{\footnotesize{Examples of variations of spectra, corresponding to the set of parameters given in Table 1.}}
\end{center}
\end{figure}

\begin{table}
\centering
\caption{\footnotesize{Parameters for Atoll type spectra.}}
\tiny{
\begin{tabular}{c c c c c c c c c}
\hline
\hline
No. & Class & $\dot{m}_d$ & $\dot{m}_h$ & $X_s$ & $T_{NBOL}$ & $f_{ic}$ & $f_{CB}$ \\~~\\
\hline
S7 & IS & 0.5 & 0.5 & 45.0 & 0.20 & 0.95 & 0.500 \\
S8 & LLB & 1.0 & 1.5 & 15.0 & 1.00 & 0.25 & 0.125 \\
S9 & LB & 0.5 & 0.5 & 30.0 & 0.60 & 0.95 & 0.500 \\
S10 & UB & 2.0 & 2.5 & 15.0 & 1.00 & 0.05 & 0.125 \\
\hline
\hline
\end{tabular}
}
\end{table}

\begin{figure}[h]
\begin{center}
\includegraphics[width=4.5cm, height=4.5cm]{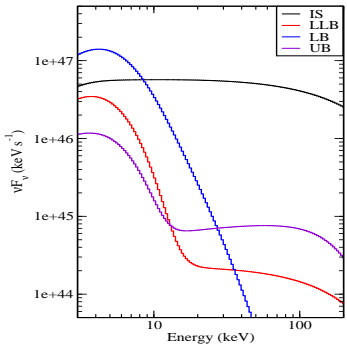}
	\caption{\footnotesize{Examples of variations of spectra, corresponding to the set of parameters given in Table 2.}} 
\end{center}
\end{figure}
\subsection{Generation of Z and Atoll tracks}
In Fig. 4, we present colour-colour diagrams of the same cases to show that the track characteristics of the transitions observed in Z and atoll sources can also be reproduced by us. These tracks in CCD can be compared with the tracks of observed X-ray spectra of Sco X-1 and 4U 1705-44 (van der Klis 1989). Fig. 2 and Fig. 4 show that the model-generated spectra captures the $HB\rightarrow NB \rightarrow FB$ transitions in 3.0-200.0 keV energy range for a Z type source and traces out the `Z' in the CCD. In Figs. 3 and 4, model generated spectra shows the $IS\rightarrow LLB\rightarrow LB \rightarrow UB$ transitions in the same energy range. The distinct Island state and Banana tracks are seen in Fig. 4 for the set of parameters used in Table 2. For consistency, the normalization (N) is kept constant for all simulated spectra. Both HC and SC were scaled by a multiplicative factor of 5 for S1 to S5 to show variations in the same plot, in Fig. 4. It can be seen by comparing S5 and S6 with S5.2 and S5.3 (Table 1), that the parameters reported here have a complex effect and in no way uniquely represent a particular state in spectra or the CCD.

\begin{figure}[h]
\begin{center}
\includegraphics[width=4.5cm, height=4.5cm]{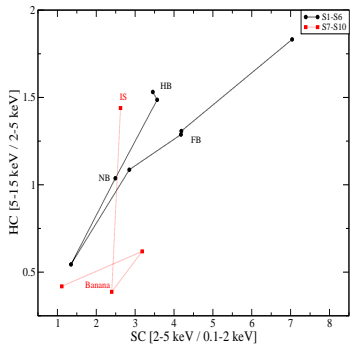}
	\caption{\footnotesize{Hard Colour (HC) vs. Soft Colour (SC) plots for the simulated spectra which follow typical tracks characteristics of the transitions found in Z and Atoll sources (see, Fig. 1, Panel 2 and Fig. 3, Panel 3 of van der Klis 1989).}}
\end{center}
\end{figure}
\subsection{X-Ray Spectral Fits}
In Fig. 5, we attempt to fit manually a few observed spectra in the $3.0-60.0$ keV energy range without adding the absorption properties at low energies. For this purpose we use the standard product data of RXTE/PCA, obtained through NASA's XAMIN archives and using the HEASOFT package under HEASARC. We obtained reasonable agreement between the theoretical and the observed spectra, especially in the $5.0-60.0$ keV range. Below 5 keV, the deviation is due to non-inclusion of absorption. The fit parameters are noted in Table 3.
Here, we presented examples of applications of the spectral model TANS for accreting Neutron Stars. We showed that with only seven independent parameters, we can reproduce spectra of all the observed states and also generate the tracks for the transitions between different branches of Z and Atoll sources. Though our code is yet to be integrated with a spectral fit software, we find encouraging signs that observed spectra of several sources could be fitted manually with reasonable accuracy with these seven parameter model alone. In all the cases we do not change the normalization as our spectra from theoretical consideration with self-consistent Comptonization and reflection are generated as a whole across the states and the normalization is needed only to scale these theoretical spectra with the observed spectra. We also report three manually fitted spectra of an IS, a LB and a HB spectrum of two different objects to impress excellent fits are possible. The deviations of the fits at lower energies are primarily due to non-inclusion of the absorption due to inter-stellar medium ($< 5 keV$). 
\begin{figure}[h]
\begin{center}
\includegraphics[width=4.5cm, height=4.5cm]{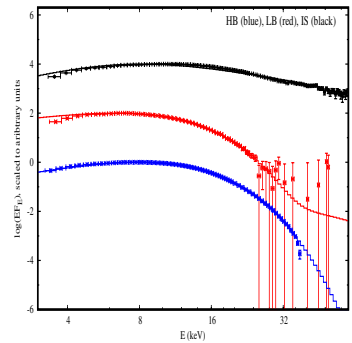}
\caption{\footnotesize{Approximate spectral fitting using the TANS model for the RXTE/PCA spectra for the HB of Sco X-1 (blue), IS of 4U 1705-44 (black), and LB of 4U 1705-44 (red). The source data have been plotted in 3.0-60.0 keV energy range, with error-bars. The corresponding solid lines show the model generated spectra to indicate fits of varied spectra are possible by changing only a few flow parameters.}} 
\end{center}
\end{figure}
\begin{table}
\centering
\caption{\footnotesize{Spectral fitting parameters of TANS model for typical HB of Sco X-1 (Obs ID: 20053-01-01-00), IS and LB state of 4U 1705-44 (Obs ID: 40034-01-09-00$^a$ and 40034-01-02-09$^b$, respectively).}}
\tiny{
\begin{tabular}{c c c c c c c c}
\hline
\hline
Source & $\dot{m}_d$ & $\dot{m}_h$ & $X_s$ & $T_{NBOL}$ & $f_{ic}$ & $f_{CB}$ \\~~\\
\hline
Sco X-1 & 1.5 & 1.0 & 10.0 & 2.10 & 0.65 & 0.325 \\
$^a$4U 1705-44 & 0.5 & 2.5 & 15.0 & 0.60 & 0.925 & 0.950 \\
$^b$4U 1705-44 & 1.5 & 2.5 & 15.0 & 1.80 & 0.25 & 0.425 \\
\hline
\hline
\end{tabular}
}
\end{table}
\section{Discussion}
The model COMPTB (Farinelli \& Titarchuk 2011) considers two separate and independent Compton clouds to successfully fit the spectra of multiple Z and Atoll sources (Titarchuk, Seifina \& Shrader 2014; Seifina et al. 2015). In our model, the geometry is based on hydrodynamic solutions and thus removes the need for separate Compton clouds to obtain the spectra, which also reduces the number of parameters to seven, as compared to 10+ in the case of COMPTB. We had previously discussed the geometry obtained from TCAF for NS and the COMPTB geometry in Bhattacharjee \& Chakrabarti 2017. In future, we would like to carry out a comparative study of spectral fitting using TANS and COMPTB. For the TANS model, we have assumed the neutron star to emit radiation in a spherically symmetric manner, with a uniform temperature $T_{NBOL}$. The spread of NBOL over the surface and the distribution of temperature of NBOL, depends on the complex inner boundary conditions of the star as well as the time-dependent behaviour of the accreting matter (for KD reaching NS: Inogamov \& Sunyaev 1999; for SKD reaching NS: Bhattacharjee \& Chakrabarti 2017, 2019), both of which are beyond the scope of this model. To account for the variation of spread of the boundary layer on the surface of the star and the consequent variation of intercepted fraction of photons from NBOL by CENBOL or vice-versa (both are proportional to the emitting area), we have introduced the factor $f_{CB}$ as a free parameter. Similarly, the density distribution within the Comptonizing region is also time-dependent and grows rapidly very close to the surface, around the density jump or inner shock (Bhattacharjee and Chakrabarti 2019). We also know that the number of scattering a photon undergoes, $n_{scat}\approx max[\tau,\tau^2]$, which strongly depends on the distribution of optical depth. As a result, most of the scattering required to produce the Comptonized spectra can be limited to a region very close to the NBOL, depending on the optical depth $\tau$, as measured from the surface of the star. For a strong inner shock (or density jump) with high accretion rates, the density close to NBOL can be high enough where $n_{scat}$ ($\sim\tau^2$) will saturate close to NBOL. On the opposite end, for weak density jumps near NBOL and lower accretion rates, the optical depth of the cloud can be low enough to allow NBOL photons to scatter throughout the cloud, as $n_{scat}\sim\tau$ and $\tau_0$ is low. In the model, we include these and other intermediate cases, by introducing the free parameter, transparency factor, $f_{ic}$. We want to conduct further detailed spectral analysis of multiple Z and Atoll sources to study the behavior of the parameters $f_{CB}$ and $f_{ic}$, in near future, to modify the model or include more parameters, if needed. A detailed study of the variation of the computed spectra with all the flow parameters will be reported soon. 
\acknowledgments
AB acknowledges the computational support provided by SNBNCBS. SKC acknowledges support from the DST/SERB sponsored Extra Mural Research project (EMR/2016/003918) fund.

\end{document}